# Reflective, polarizing, and magnetically soft amorphous Fe/Si multilayer neutron optics with isotope-enriched $^{11}B_4C$ inducing atomically flat interfaces

A. Zubayer,[1,*] N. Ghafoor,[1] K. A. Thórarinsdóttir,[2] S. Stendahl,[1] A. Glavic,[3] J. Stahn,[3] G. Nagy,[4] G. Greczynski,[1] M. Schwartzkopf,[5] A. Le Febvrier,[1] P. Eklund,[1,*] J. Birch,[1,*] F. Magnus,[2,*] F. Eriksson[1,*]

1. Thin Film Physics Division, Department of Physics, Chemistry and Biology (IFM), Linköping University, SE-581 83 Linköping, Sweden
2. Science Institute, University of Iceland, Dunhaga 3, IS-107 Reykjavik, Iceland
3. Paul Scherrer Institut, 5232 Villigen PSI, Switzerland
4. Department of Physics and Astronomy, Uppsala University, SE-75120, Uppsala, Sweden
5. Photon Science, DESY, Notkestraße 85, 22607, Hamburg, Germany

*   E-mail: anton.zubayer@liu.se ; per.eklund@liu.se ; jens.birch@liu.se ; fridrikm@hi.is ; fredrik.eriksson@liu.se

**The utilization of polarized neutrons is of great importance in scientific disciplines spanning materials science, physics, biology, and chemistry. Polarization analysis offers insights into otherwise unattainable sample information such as magnetic domains[1] and structures,[2] protein crystallography,[3] composition, orientation, ion-diffusion mechanisms,[4] and relative location of molecules in multicomponent biological systems.[5]. State-of-the-art multilayer polarizing neutron optics have limitations, particularly low specular reflectivity and polarization at higher scattering vectors/angles, and the requirement of high external magnetic fields to saturate the polarizer magnetization. Here, we show that by incorporating $^{11}B_4C$ into Fe/Si multilayers, amorphization and smooth interfaces can be achieved, yielding higher neutron reflectivity, less diffuse scattering and higher polarization. Magnetic coercivity is eliminated, and magnetic saturation can be reached at low external fields (>2 mT). This approach offers prospects for significant improvement in polarizing neutron optics, enabling; nonintrusive positioning of the polarizer, enhanced flux, increased data accuracy, and further polarizing/analyzing methods at neutron scattering facilities.**

The importance of polarized neutrons at the forefront of today's materials science, physics, chemistry, and biology is underscored by their broad applicability for characterization of thin films, nanostructures, powders, liquids, and crystals. These investigations provide insights into crystal structures, magnetic properties, diffusion, film roughness/intermixing and phonon dispersions.[6] Nevertheless, inherent limitations to state-of-the art neutron optics limit their broad applicability. Specifically, there is a need for improvement in four key areas: reflective capabilities (higher intensity and reflectivity at higher scattering vectors/angles), polarization (enhanced polarization at low scattering angles/vectors and the possibility for polarization at higher scattering angles/vectors), reduction of magnetic coercivity, and reduced diffuse scattering. Here, we address all four of these challenges by incorporation of low-neutron-absorbing isotope-enriched $^{11}B_4C$ in Fe/Si polarizing multilayer neutron optics deposited by DC-magnetron sputtering.

First, to enhance reflectivity, it is necessary to reduce the interface width between the layers in the multilayer structure. Currently, the interface width (and/or iron silicide) of Fe/Si polarizing multilayer neutron optics is usually around 8-11 Å.[7] By preventing silicide formation and minimizing roughness, the reflectivity can be significantly increased. In addition, thinner layers can be deposited, allowing reflection at higher scattering vectors.

Second, achieving high polarization requires a suitable contrast or match between the nuclear and magnetic scattering length densities (SLD) of the magnetic and non-magnetic layers, depending on the spin state. The sum of the nuclear SLD and the magnetic SLD of the magnetic layer should exhibit high contrast with the SLD of the non-magnetic layer. Conversely, the difference between the nuclear SLD and the magnetic SLD of the magnetic layer should match the SLD of the non-magnetic layer. This arrangement ensures that spin-up neutrons are reflected while spin-down neutrons are not, thereby achieving high polarization.[8]

Third, Fe/Si multilayers typically exhibit a crystalline structure, leading to the presence of magnetic domains and a high magnetic coercivity, which is typically associated with magnetic particle/grain size.[9] The high coercivity can be addressed by the use of magnetically soft materials, i.e., materials that can be easily magnetized and demagnetized, yielding low magnetic coercivity. For example, the use of soft Fe can reduce the magnetic coercivity by several orders of magnitude.[10] By amorphizing the Fe layers,[11] a smaller external field would be required to saturate the magnetization, allowing the optics to be positioned very close to the sample environment to enable polarization without magnetic stray fields affecting the sample environment itself.

Fourth, it is crucial to minimize and characterize the diffuse scattering from neutron optical elements to be able to deduce whether the scattered neutrons come from the polarizer or the sample of interest. Since diffuse scattering from multilayer optics originates from interfacial roughness as well as nanocrystallites, it will be largely reduced by amorphization and formation of flatter interfaces. We examine the diffuse scattering outside the specular view so that lateral correlations[12] in the multilayers can be quantified.

The incorporation of isotope-enriched $^{11}B_4C$ presents an opportunity to address these issues. $^{11}B_4C$, instead of just $^{11}B$, is used since $^{11}B_4C$ sputter targets are more stable, cheaper, and available. First, $^{11}B_4C$ could reduce interface roughness[13] through elimination of crystalline grain growth by amorphization and prevent silicide formation through the strong bonds between B and Fe. Second, the high SLD of $^{11}B_4C$ allows for tuning the SLDs. Third, by eliminating lateral correlations, diffuse scattering can be minimized. Fourth, amorphization also eliminates magnetic coercivity. The amorphization is due to the stronger Fe-B bonds than Fe-Fe, or Fe-Si bonds, where the enthalpies of formations are -38 and -26 kJ/mole for Fe-B and Fe-Si, respectively.[14] Consequently, preventing the formation of iron-silicides and crystalline Fe leads to a reduction in interface width. The concept of SLD tuning using different amounts of $^{11}B_4C$ incorporated in the Fe and Si layers, is illustrated in Figure 1 (a-b).

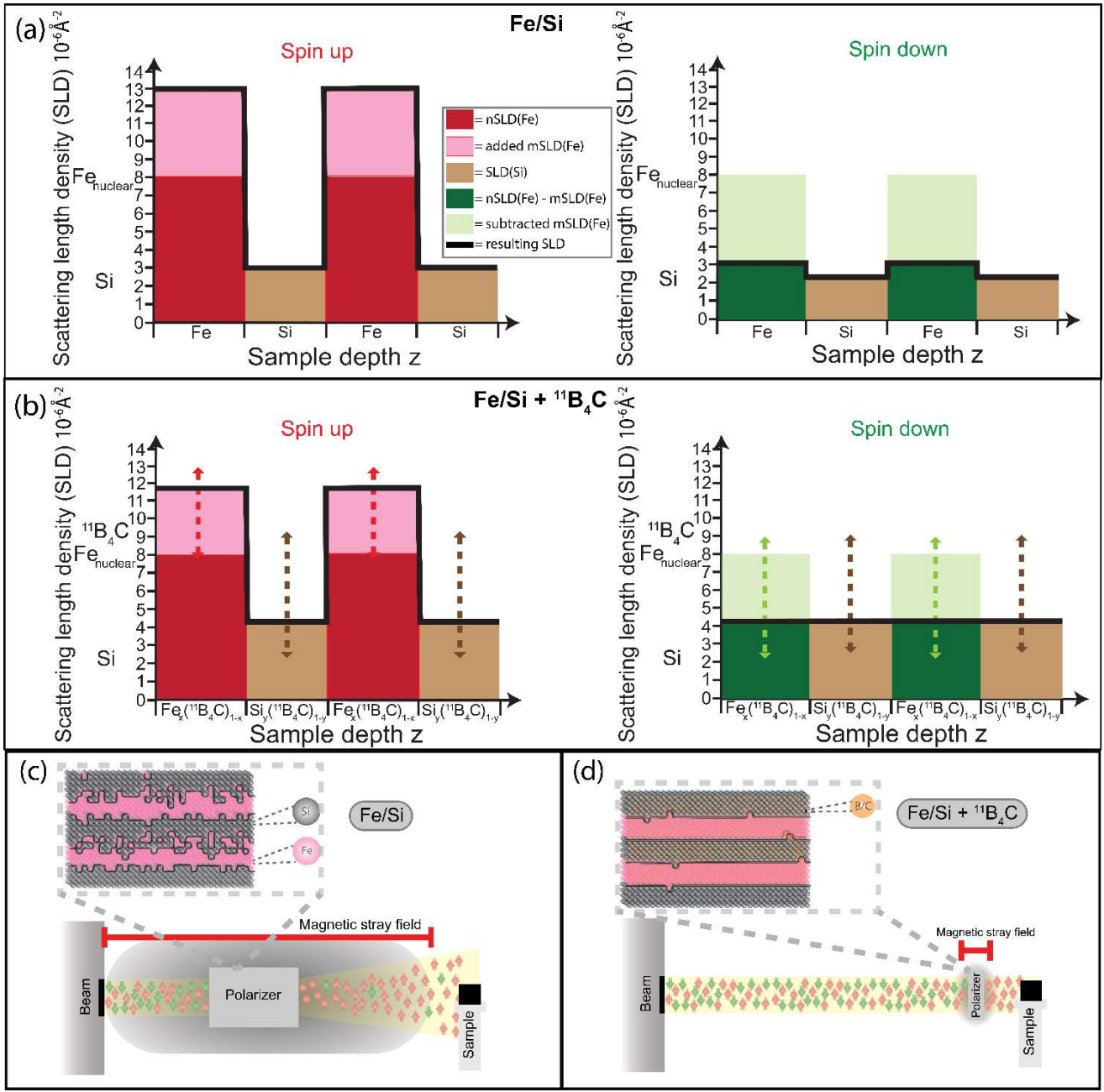

**Figure 1. Concept of scattering length density (SLD) tuning and implementation of improved polarizing neutron optics.** (a, b) Illustrations showing the SLD depth profiles (solid black curves) resulting from the different magnetic and nuclear SLDs for (a) Fe/Si and (b) $^{11}B_4C$-containing Fe/Si multilayers. In the spin-up configuration, the magnetic SLD (mSLD) of Fe is added to the nuclear SLD (nSLD) of Fe, while in the spin-down configuration, the magnetic SLD of Fe is subtracted from the nuclear SLD of Fe. The dashed arrows in (b) represent the SLD tuning possibilities achieved by incorporating varying amounts of $^{11}B_4C$. (c-d) Comparison of Fe/Si (c) with Fe/Si + $^{11}B_4C$ (d) highlighting the advantage of utilizing a lower applied magnetic field for polarizer saturation. This, in turn, allows for nonintrusive positioning of the polarizer closer to the sample environment, which can be achieved by incorporating $^{11}B_4C$ to eliminate magnetic coercivity, in other words, make the polarizer magnetically soft.

Figure 1(a-b) illustrates the SLD depth profiles for spin-up and spin-down configurations for (a) Fe/Si and (b) Fe/Si + $^{11}B_4C$, i.e., with $^{11}B_4C$ incorporated in both the Fe and the Si layers in varying amounts.

The goal of polarization is to maximize contrast for spin-up and minimize it for spin-down. The Fe/Si configuration, with some contrast in spin-down, can be improved by adding $^{11}B_4C$ to the Si layer to increase its SLD. However, diluting Fe with $^{11}B_4C$ decreases the magnetic SLD (mSLD) of Fe, which can be compensated by further increasing the concentration of $^{11}B_4C$ in Si. The dashed arrows indicate the range of tunability via $^{11}B_4C$ incorporation at different concentrations. SLD tuning has been attempted using ion-beam sputtering and Ar implantation during Fe/Ge multilayer growth,[15] and by the addition of $N_2$ and $O_2$ reactive gases during growth of Fe/Si supermirrors.[16] However, these methods were limited by attainable SLDs and challenges in trapping gas elements. The current approach allows larger SLD tuning (from $2.1 \cdot 10^{-6}$ to $9.0 \cdot 10^{-6}$ Å$^{-2}$), which is much wider than previous methods, leading to maximum polarization and precise layer SLD adjustment in a thin film. A comparison between Fe/Si (Figure 1(c)) with Fe/Si + $^{11}B_4C$ (Figure 1(d)) indicates the effect of a reduction in magnetic stray field owing to a lower applied field needed to reach magnetic saturation for a magnetically soft polarizer due to amorphization of Fe,[11] in accordance with the results presented in this work. Additionally, a comparison of the two configurations illustrates the decrease in diffuse scattering for Fe/Si + $^{11}B_4C$, attributed to the reduction of lateral correlations that give rise to the diffuse scattering in pure Fe/Si.

Samples of Fe/Si and Fe/Si + $^{11}B_4C$ were grown using magnetron sputter deposition and were studied using X-ray diffraction (XRD), X-ray reflectivity (XRR), grazing incidence small-angle X-ray scattering (GISAXS) and grazing incidence wide-angle scattering (GIWAXS), elastic recoil detection analysis (ERDA), polarized neutron reflectivity (PNR), vibrating sample magnetometry (VSM), transmission electron microscopy (TEM) and X-ray photoelectron spectroscopy (XPS). Throughout this study we refer to N as the number of bilayers in the multilayer. Γ as the layer thickness ratio being the thickness of the Fe layer divided by the bilayer thickness (i.e. Fe layer / (Fe layer + Si layer)). Λ as the bilayer thickness.

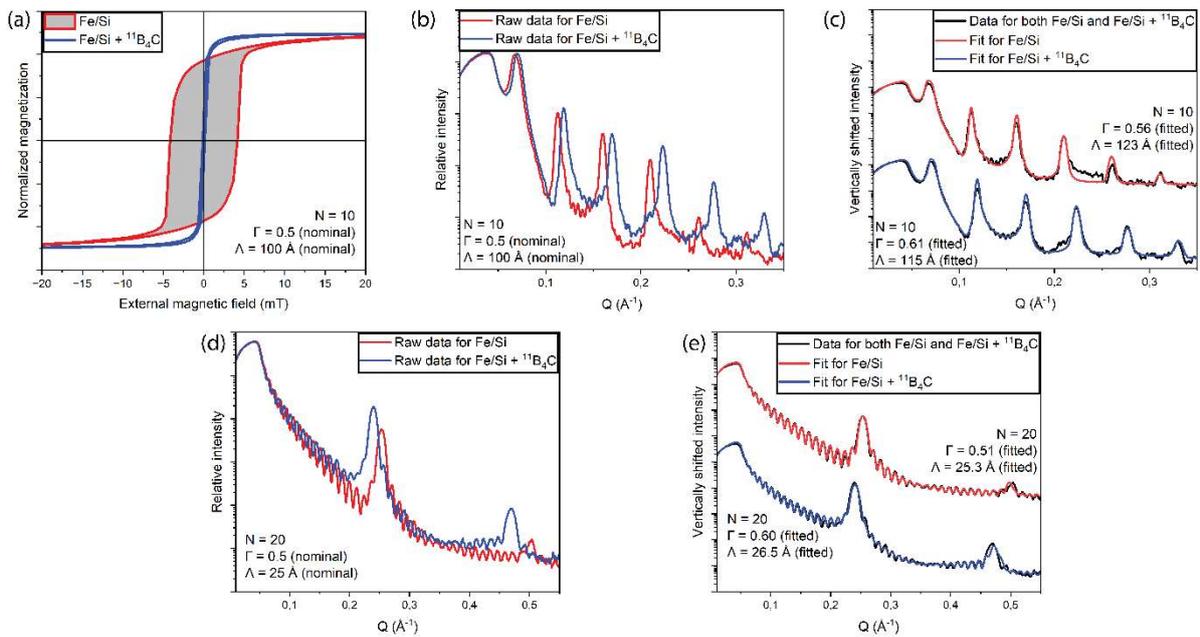

**Figure 2. Magnetism and interface improvement.** (a) Hysteresis curves showing the normalized magnetization measured by vibrating sample magnetometry for Fe/Si and Fe/Si + $^{11}B_4C$ multilayers. The grey area represents the magnetic energy that is dissipated during the magnetization reversal process within the material, which is eliminated in the case of Fe/Si + $^{11}B_4C$ multilayers. (b) X-ray reflectivity data of an Fe/Si and an Fe/Si + $^{11}B_4C$ multilayer, both with N = 10 bilayers, a nominal layer thickness ratio of $\Gamma = 0.5$ and a nominal bilayer thickness $\Lambda = 100$ Å. (c) XRR fits of the data seen in (b). The fitting parameters resulted in a 13 Å interface width for Fe/Si and a 6 Å interface width for Fe/Si + $^{11}B_4C$. (d) X-ray reflectivity data of an Fe/Si and an Fe/Si + $^{11}B_4C$ multilayer sample, both with N = 20 bilayers, a nominal layer thickness ratio of $\Gamma = 0.5$ and a nominal bilayer thickness $\Lambda = 25$ Å. (e) XRR fits of the data seen in (d). The fitting parameters showed that in the Fe/Si case, the layers had a 13 Å Fe layer and a 13 Å Fe and Si mixed layer, indicating the absence of pure Si layers. In contrast, for Fe/Si + $^{11}B_4C$, a 5.5 Å interface width was observed.

Initial series of experiments were performed to determine suitable concentrations of $^{11}B_4C$ and multilayer periods, as detailed in Supplementary Information Section S1. Based on these initial experiments, a content of 14 vol.% $^{11}B_4C$ was chosen for both Fe and Si layers to maintain the Fe layer in an X-ray amorphous state. The magnetic properties of Fe/Si and Fe/Si + $^{11}B_4C$ with 14 vol.% $^{11}B_4C$ in the Fe and Si layers were studied using VSM. The results, presented in Figure 2 (a), demonstrate that the inclusion of 14 vol.% of $^{11}B_4C$ completely eliminated the coercivity, resulting in a five-fold reduction in the external magnetic field required for magnetization saturation. The Fe/Si sample required approximately 10 mT for saturation, whereas Fe/Si + $^{11}B_4C$ only required approximately 2 mT. Figure 2 (b-c) displays the XRR measurements and fits for Fe/Si and Fe/Si + $^{11}B_4C$ multilayers with a nominal bilayer thickness of 100 Å and nominal thickness ratio of 0.5. The higher peak intensities for the Fe/Si + $^{11}B_4C$ sample is indicative of a decreased interface width compared to Fe/Si multilayer. Furthermore, fits of the XRR measurements for Fe/Si and Fe/Si + $^{11}B_4C$ multilayers with a nominal bilayer thickness of 25 Å in Figure 2 (e) indicate that the Fe/Si sample, although there is a chemical modulation, exhibits extensive mixing of Fe and Si throughout the entire sample, likely due to the expected formation of iron silicide. In contrast, the addition of $^{11}B_4C$ prevents the formation of iron silicide and other crystalline phases, as indicated by Figure S1(a). The interface width, obtained through the fitting, for Fe/Si + $^{11}B_4C$ was 4.8 Å, and thus the incorporation of $^{11}B_4C$ leads to more distinct separation between magnetic and non-magnetic layers and reduction in interface width. Regardless of bilayer thickness, the inclusion of $^{11}B_4C$ resulted in decreased interface width and significantly increased reflectivity.

An extended range of samples was grown using ion-assisted magnetron sputter deposition in a different chamber and characterized (see Supplementary Section S1). The results obtained from this extended study corroborate the presented findings and confirm the robustness and reproducibility of the observed effects.

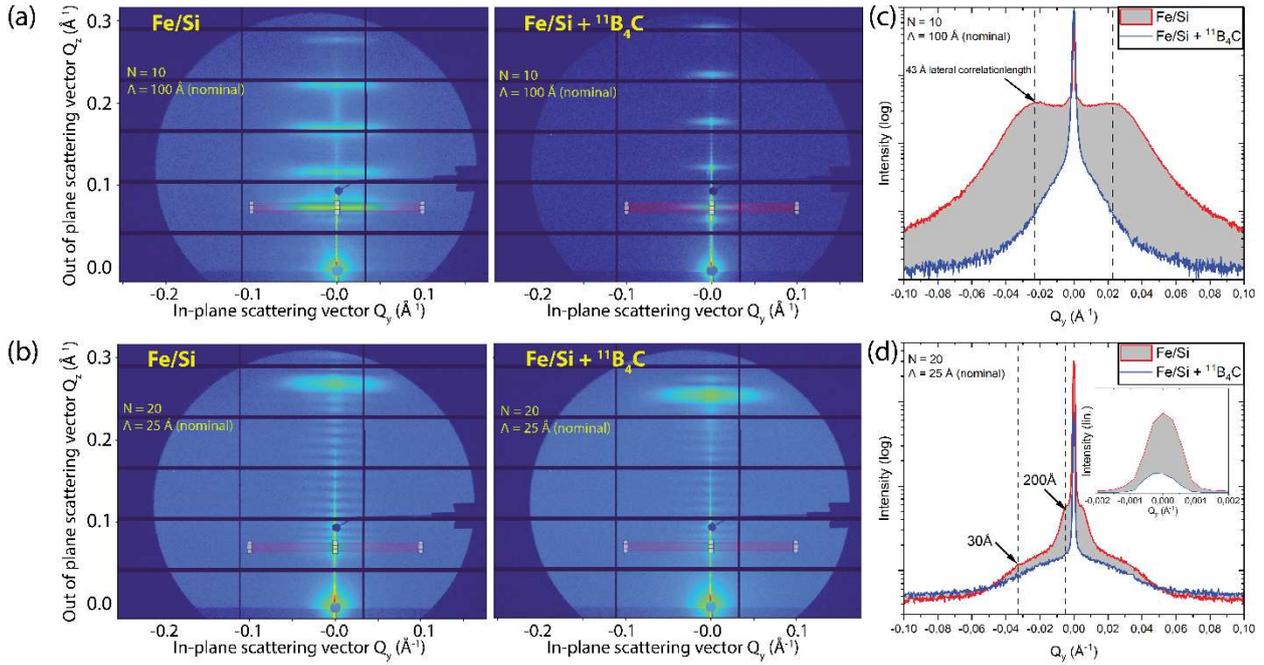

**Figure 3. Lateral correlation and diffuse scattering analysis.** Grazing-incidence small-angle X-ray scattering (GISAXS) measurements were performed on Fe/Si and Fe/Si + $^{11}B_4C$ samples. (a) and (b) present the raw GISAXS maps for Fe/Si and Fe/Si + $^{11}B_4C$, respectively, for bilayer thicknesses of $\Lambda$ = 100 Å and 25 Å and number of bilayers N = 10 and 20, respectively, and a layer thickness ratio of $\Gamma$ = 0.5. The red rectangles highlight the first off-specular Bragg-sheet. (c-d) shows the horizontal line scan of the first Bragg sheet of Fe/Si (black) and Fe/Si + $^{11}B_4C$ (red) of 100 Å, N = 10 and 25 Å, N = 20, respectively, and a layer thickness ratio of $\Gamma$ = 0.5, derived from the GISAXS raw data seen in (a) and (b). The inset in (d) shows the intensity of the Bragg sheet in the $Q_y$-direction on a linear scale.

To investigate the off-specular/diffuse scattering and the lateral correlations,[12,17,18] GISAXS was used. The constructive interference of repeating surface and interface features in the sample were examined to assess the lateral correlations in the in-plane direction. Figure 3 (a) and (b) presents the raw GISAXS data obtained from multilayers of (a) Fe/Si and (b) Fe/Si + $^{11}B_4C$ for nominal bilayer thicknesses $\Lambda$ and number of bilayers N being $\Lambda$ = 100 Å, N = 10 in (a) and $\Lambda$ = 25 Å, N = 20 in (b). The results indicate a significant reduction in off-specular scattering in the Fe/Si + $^{11}B_4C$ sample compared to the Fe/Si sample, highlighting the impact of $^{11}B_4C$ incorporation on mitigating lateral correlations. To quantitatively deduce the lateral correlations causing the off-specular scattering, line scans were taken of the first Bragg sheet, as indicated by the red rectangles in Figure 3 (a) and (b). These line scans show the intensity distribution in the $Q_y$-direction. The grey area between the black (Fe/Si) and red (Fe/Si + $^{11}B_4C$) lines, in Figure 3(c-d) represents the reduction in diffusely scattered intensity at the first Bragg sheet in the Fe/Si sample compared to the Fe/Si + $^{11}B_4C$ sample, showing a significant reduction in diffuse scattering in the presence of $^{11}B_4C$.

The Fe/Si samples, seen in Figure 3(c-d), exhibits shoulders, corresponding to constructive interference of repeating features in the lateral direction. The obtained lateral correlation length is approximately 43 Å, as determined from the position of the shoulders in Figure 3(c). However, when $^{11}B_4C$ is incorporated, these periodic correlations are eliminated. This can be attributed to the elimination or significant reduction in crystallite formation caused by the incorporation of $^{11}B_4C$. Hence the lateral correlations associated with crystalline Fe or iron silicides are also diminished as seen in Figure 3. In Figure 3(d), the line scan of the Fe/Si + $^{11}B_4C$ samples with $\Lambda = 25$ Å and N = 20 is shown and it is observed that the incorporation of $^{11}B_4C$ reduces the off-specular scattering and eliminates the shoulders. The presence of shoulders in the Fe/Si sample suggests the presence of larger hills with smaller hills on top in the lateral directions, with lateral correlation lengths of approximately 200 Å and 30 Å. The inset in Figure 3(d) shows the linear intensity of the off-specular scattering in the $Q_y$-direction. It clearly illustrates that the Fe/Si sample exhibits a higher off-specular intensity compared to Fe/Si + $^{11}B_4C$. This further emphasizes the importance of incorporating $^{11}B_4C$ to reduce the diffuse scattering of the reflected beam and improve the overall sample characteristics.

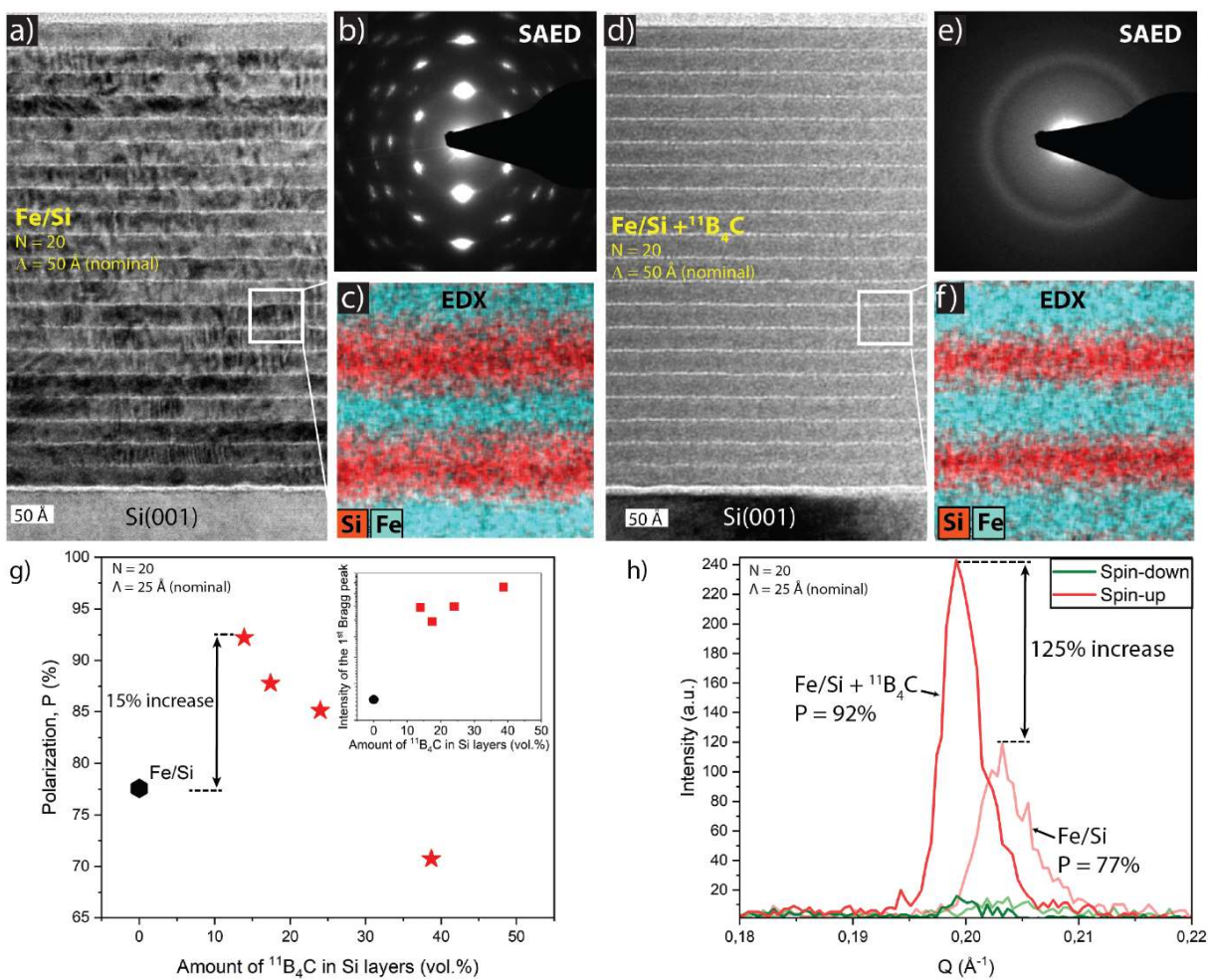

**Figure 4. Microscopy, polarization and reflectivity.** (a-c) Overview diffraction contrast cross-sectional TEM micrograph, corresponding selected area electron diffraction (SAED) pattern, and colored coded (Fe-cyan, Si-red) 135 x 135 Å$^2$ energy dispersive X-ray (EDX) elemental map of Fe/Si multilayer, respectively. (d-f) Overview diffraction contrast TEM micrographs, corresponding SAED pattern, and colored coded (Fe-green, Si-red) 135 x135Å$^2$ EDX elemental map of Fe/Si + $^{11}B_4C$ multilayer with 14 vol.% of $^{11}B_4C$, respectively. Both multilayers have bilayer thickness of $\Lambda = 50$ Å and N = 20 number of bilayers. (g) Polarization (P) at the Bragg peak for four samples, indicated with red stars, with different ratios between $^{11}B_4C$ and Si. The polarization of the Fe/Si sample is represented by the black hexagon. The inset shows the intensity of the first Bragg peak for the spin-up case for these samples. (h) Polarized neutron reflectivity in the Q-range between 0.18 to 0.22 Å$^{-1}$ to visualize the 1$^{st}$ Bragg peak, for Fe/Si and the best polarizing Fe/Si + $^{11}B_4C$ sample. The polarization values are obtained through 10-minute measurements at the specific Bragg peak positions.

Comparing TEM and electron diffraction analyses, as seen in Figure 4(a-b) for Fe/Si and Figure (d-e) for Fe/Si + $^{11}B_4C$, suggests that adding $^{11}B_4C$ prevents iron silicide formation and leads to an amorphous multilayer, also in good agreement with XRD in S1(a). GIWAXS measurements, in S4, also support this, showing the 14 vol.% $^{11}B_4C$ sample as nearly fully amorphous, while Fe/Si shows clear crystallinity. The images in 4(a) and 4(d) appear as if the Fe layer is very thick compared to the Si layer, this could be due to an instrumental artifact. Note that the Fe and Si layers are not clearly separated and there seems always to be a gradient of Fe and Si content throughout the multilayer for both samples. However, comparing the EDX maps of Figure 4(c) and Figure 4(f), the intermixing between Fe and Si atoms is much smaller when $^{11}B_4C$ is incorporated.

Polarized neutron reflectivity (PNR) measurements, as seen in Figure 4(g), aim not only to validate enhanced reflectivity, polarization, and SLD tunability, but also to demonstrate these achievements at higher scattering angles/vectors than typical polarizing neutron optics. State-of-the-art commercial optics can reach reach reflectivity up to $Q = 0.119$ Å$^{-1}$.[19] PNR measurements show better polarization for certain $^{11}B_4C$ concentrations in Si compared to Fe/Si. The optimal sample provides 15% higher polarization and 125% higher reflectivity, demonstrating the benefits of $^{11}B_4C$ in SLD matching and reducing interface width for higher reflectivity. To explore the impact for thicker layers, reflectivity and polarization were simulated using the parameters obtained from the reflectivity fits in Figure 2 (b) or (c) for $\Lambda = 200$ Å and $N = 7$ (Supplementary Information Section S5). These results showed that the reflectivity difference is not as pronounced for thicker bilayers as for thinner ones, as seen in Figure 4(d), and the polarization increase is only 1% compared to the 15% observed for thinner bilayers.

In summary, this study has addressed the four main outstanding issues concerning polarizing multilayer neutron optics. First, the incorporation of $^{11}B_4C$ in Fe/Si multilayers leads to higher reflectivity, which can enhance the neutron flux. Second, the improved polarization through SLD tuning with the inclusion of $^{11}B_4C$ enables experiments that require high polarization sensitivity. Third, the reduction in diffuse scattering in the polarizer is beneficial for experiments involving the study of diffuse scattering from the sample. Finally, the elimination of coercivity allows for the polarizer to be positioned closer to the sample environment. This yields greatly improved reflectivity and polarization, and offers prospects for experiments with nonintrusive positioning of the polarizer near the sample environment with enhanced flux, data accuracy, and polarizing/analyzing methods in neutron scattering facilities.

**ADDITIONAL INFORMATION**

Supplementary information is available in the online version of the paper. Correspondence and requests for materials should be addressed to A.Z., P.E., J.B., F.M., and F.E.

**Data availability statement**

Essential data generated or analyzed during this study are included in this published article (and its supplementary information files). The datasets generated during and/or analyzed during the current study will be made available after acceptance in the Zenodo or other repository at DOI.

**COMPETING FINANCIAL INTERESTS**

The authors declare no competing financial interests.

## AUTHOR CONTRIBUTIONS

A.Z., N.G., J.B., and F.E. conceived and initiated the work. F.E. supervised the work.

A.Z. planned the experiments, grew the samples, characterized (XRR, XRD, GISAXS, PNR) and analyzed them through fitting (GenX 3) and interpretation.

N.G. performed the TEM.

K.A.Th. and F.M. performed and analyzed the VSM.

A.G. and J.S. contributed to the PNR and interpretation.

G.N. performed and analyzed the ERDA.

G.G. performed and analyzed the XPS measurements.

S.S. and M.S. contributed to the GISAXS and GIWAXS and interpretation.

A.l.F and P.E. contributed to the planning and implementation of the film growth and analysis.

A.Z., N.G., P.E. and F.E wrote the paper. All coauthors read and commented on successive versions of the manuscript.

**Acknowledgments**

The authors acknowledge the Swedish Government Strategic Research Area in Materials Science on Functional Materials at Linköping University (Faculty Grant SFO-Mat-LiU No. 2009 00971). We thank Paul Scherrer Institute (PSI), Switzerland, for several beamtimes at MORPHEUS. We acknowledge the Swedish Research Council (VR) under the project number 2019-04837_VR (F.E.) and 2018-05190_VR (N.G). A.Z. acknowledge the support from Hans Werthéns grant (2022-D-03) and the Royal Academy of Sciences Physics grant (PH2022-0029). P.E. and A.l.F. acknowledge the Knut and Alice Wallenberg foundation through the Wallenberg Academy Fellows program (KAW-2020.0196) and the Swedish Research Council (VR) under project number 2021-03826. K.A.Th. and F.M. acknowledge funding from the Icelandic Research Fund (grant no. 217843-053). Parts of this research were carried out at the light source PETRA III at DESY, a member of the Helmholtz Association (HGF). Accelerator operation, in Uppsala, is supported by the Swedish Research Council VR-RFI (Contracts No. 2017-00646_9 and 2019-00191) and the Swedish Foundation for Strategic Research (Contract No. RIF14-0053). Justinas Palisaitis is acknowledged for performing STEM/EELS analysis and Swedish Research Council and SSF for access to ARTEMI, the Swedish National Infrastructure in Advanced Electron Microscopy (2021-00171 and RIF21-0026).

**Methods**

Fe/Si and Fe/Si + $^{11}B_4C$ multilayer thin films were deposited using DC magnetron sputter deposition in an ultra-high vacuum system with a background pressure of about $2.6 \cdot 10^{-7}$ Pa ($2 \cdot 10^{-9}$ Torr). Further details of the deposition system can be found elsewhere.[20] The multilayers were deposited on 10x10x0.5 mm$^3$ Si(100) substrates with a native oxide. The substrate was kept under constant rotation of 15 rpm and not intentionally heating was used during the deposition. 50 mm targets of Fe (Plasmaterials 99.95% purity), $^{11}B_4C$ (RHP Technology, 99.8% chemical purity, >90% isotopic purity), and Si (Kurt J. Lesker 99.95% purity) were used and powered at 33 W, 50 W and 10-40 W, respectively. The sputtering rates were determined from X-ray reflectivity thickness measurements of multilayers. The multilayers were achieved by using computer-controlled shutters where the opening durations were selected for the desired layer thicknesses and design of the multilayer. During the deposition, a substrate bias of -30 V was applied to the substrate. When depositing Fe/Si + $^{11}B_4C$, each bilayer consisted of co-sputtered $^{11}B_4C$ with Fe, followed by the deposition of $^{11}B_4C$ with Si. The deposition rates of both Fe + $^{11}B_4C$ and Si + $^{11}B_4C$ were nearly equal, approximately 0.5 Å/s. When preparing samples with different ratios of Si to $^{11}B_4C$, only the deposition rate of Si was adjusted by tuning its target power.

Hard X-ray reflectivity analysis was carried out using an Empyrean diffractometer by Panalytical in a parallel beam geometry with a line-focused copper anode source operating at 45 kV and 40 mA giving Cu-K$_a$ radiation with a wavelength of 1.54 Å. In the incident beam path, a parabolic X-ray mirror was used along with a ½° divergence slit to condition the beam and limit the X-ray spot size on the sample. In the diffracted beam path, a parallel plate collimator with a parallel plate collimator slit (0.27°) was used, followed by a PIXcel detector operating in open detector mode.

X-ray diffraction (XRD) was performed using a Panalytical X'Pert diffractometer in Bragg-Brentano geometry with a Bragg-Brentano HD incident beam optics module with a ½° divergence slit and a ½° anti-scatter slit. On the secondary optics side, a 5 mm anti-scatter slit was used together with an X'celerator detector operating in scanning line mode. Diffraction measurements were performed using a 2θ scanning range of 20° to 90°, a step size of 0.018° and a collection time of 20 s.

To analyze the X-ray reflectivity (XRR) data, the multilayer bilayer thickness and interface roughness were determined through fitting. The fitting process involved the software GenX (v3),[21] which enabled the simulation of the ratio between $^{11}B_4C$ and Si in the non-magnetic layer in order to achieve the necessary SLD match between the magnetic and non-magnetic materials. Additionally, GenX was used to fit the retrieved data from corresponding X-ray and neutron reflectivity measurements.

The magnetic properties of the samples were analyzed using vibrating sample magnetometry (VSM) in a longitudinal geometry at room temperature. Magnetic hysteresis curves were measured in a field range of -20 mT to 20 mT, providing information on the saturation magnetization and coercive field of the samples.

Time-of-Flight Elastic Recoil Detection Analysis (ToF-ERDA) at the Tandem Laboratory of Uppsala University[22] was used to determine the elemental composition of selected samples. For the ToF-ERDA measurements, a beam of 36 MeV $I^{8+}$ ions was used at an incident angle of 67.5° with respect to the surface normal and the time-of-flight and energy detectors were placed at a 45° angle with respect to the incident beam direction. The Potku code[23] was used for data analysis. ToF-ERDA is a depth resolved and quantitative technique to determine sample composition, without any matrix effects or need to use reference standards. In this case the depth resolution is not sufficient to distinguish the individual layers in the multilayer samples, which are on the order of a few Ångström. Therefore, the simulated and calculated composition is an effective value, averaged over the layers.

The polarized neutron reflectivity (PNR) experiments were conducted using the polarized neutron reflectometer MORPHEUS, located at SINQ (Paul Scherrer Institut, Villigen, Switzerland). In PNR experiments, a non-polarized neutron beam initially encounters the polarizer, which polarizes the beam to allow only one spin orientation to pass through. The spin state (up or down) can be switched by a spin flipper. Subsequently, the polarized neutron beam is directed at the sample surface at a small angle (θ) and undergoes specular reflection. The reflected intensity is determined by the reflection and transmission at all interfaces. PNR provides sensitivity to the spin-dependent scattering length density (SLD) of the sample, enabling the investigation of its magnetic properties. The two possible spin orientations result in to two distinct PNR reflectivity curves. The observed Bragg peaks arise from constructive interference. The samples were measured in an external magnetic field that was sufficiently strong to magnetically saturate them in the in-plane direction, with an approximate field strength of 20 mT at the sample environment. The measurements were carried out from 0 to 15° in 2θ using a wavelengthof 4.83 Å.

The preparation of cross-sectional transmission electron microscopy (TEM) samples involved mechanical grinding and polishing, followed by Ar-ion beam milling until electron transparency was achieved. The FEI Tecnai G2 TF 20 UT microscope operating at 200 kV was used to conduct TEM.

X-ray Photoelectron Spectroscopy (XPS) was used to analyze the effect of adding $^{11}B_4C$ to the Fe/Si multilayers on the Fe-Si bond formation. The equipment used was Kratos Axis Ultra DLD instrument with a monochromatic Al Kα radiation of 1486.6 eV. The pass energy was set to 160 eV for the survey scans and to 20 eV for higher energy resolution scans. In the latter case a full width at half maximum of the Ag $3d_{5/2}$ peak from reference sample was 0.55 eV. Spectrometer was calibrated by measuring Au $4f_{7/2}$, Ag $3d_{5/2}$, and Cu $2p_{3/2}$ peak positions from sputter-etched Au, Ag, and Cu samples and comparing obtained values to the recommended ISO standards for monochromatic Al Kα sources.[24] The base pressure during XPS analyses was better than $1.1 \cdot 10^{-9}$ Torr ($1.5 \cdot 10^{-7}$ Pa). The Fe 2p, Si 2p, O 1s, B 1s and C 1s core level spectra were recorded. Sputter depth profiles were done using 0.5 keV $Ar^+$ ion beam incident at an angle of 20° from the surface and rastered over an area of 3x3 $mm^2$. All spectra are charge referenced to the Fermi level.[25] The scanning area from where the spectra were collected was $0.3 \times 0.7$ $mm^2$ centered in the middle of the etched area. Two samples were studied by XPS: Fe/Si and Fe/Si + $^{11}B_4C$ multilayers, both grown on $15 \times 15$ $mm^2$ Si substrates. Both multilayers had 10 bilayers/periods and a bilayer thickness of 100 Å while the thickness ratio in each bilayer was 1:1 (Fe:Si) nominally. Spectra was taken from within the Si layer.

We conducted GISAXS experiments using the Micro- and Nanofocus X-ray Scattering beamline (MiNaXS/P03) located at the PETRA III third generation synchrotron source of the Deutsches Elektronen-Synchrotron (DESY) in Hamburg, Germany.[26] A beam of 32 × 27 μm² (H × V) shape with a monochromatic X-ray energy of 13 keV was used. For our detector, we utilized a PILATUS 2M from Dectris in Switzerland, which had a resolution of 1475 × 1679 pixels (H × V) with a pixel size of 172 μm in both the horizontal and vertical directions. The sample-to-detector distance (SDD) was set at 2470 mm. To prevent the detector from being saturated, we masked both the direct and specular reflected beam using two separate point-like beam stops. We selected $\alpha_i = 0.45°$ as the angle of incidence in the GISAXS geometry, which was greater than all critical angles of the materials involved.

## References for Methods

# Reflective, polarizing, and magnetically soft amorphous Fe/Si multilayer neutron optics with isotope-enriched $^{11}B_4C$ inducing atomically flat interfaces

A. Zubayer[1], N. Ghafoor[1], K. A. Thórarinsdóttir[2], S. Stendahl[1], A. Glavic[3], J. Stahn[3], G. Nagy[4], G. Greczynski[1], M. Schwartzkopf[5], A. Le Febvrier[1], P. Eklund[1], J. Birch[1], F. Magnus[2], F. Eriksson[1]

*1. Thin Film Physics Division, Department of Physics, Chemistry and Biology (IFM), Linköping University, SE-581 83 Linköping, Sweden*

*2. Science Institute, University of Iceland, Dunhaga 3, IS-107 Reykjavik, Iceland*

*3. Paul Scherrer Institut, 5232, Villigen PSI, Switzerland*

*4. Department of Physics and Astronomy, Uppsala University, SE-75120, Uppsala, Sweden*

*5. Photon Science, DESY, Notkestraße 85, 22607, Hamburg, Germany*

### Section S1. Determination of $^{11}B_4C$ content and amorphization conditions

To investigate the effects of $^{11}B_4C$ on Fe-layer amorphization, roughness and formation of iron silicide, X-ray diffraction (XRD) analysis was performed of samples synthesized with different concentrations of $^{11}B_4C$. These samples were grown in a different sputtering chamber, see S7. The crystalline nature of Fe/Si multilayers for 0 vol.%, 2.5 vol.% and 5 vol.% is indicated by the presence of Fe (110) and/or $Fe_3Si$ (220) peaks that combine to form a broad diffraction peak around 45° 2θ. The width of this peak is a result of the presence of grains with varying sizes.[7] 14 vol.% on the other hand was enough to yield X-ray amorphous multilayers, free from Fe or Fe-silicide crystallites. It has been established that amorphous ferromagnets exhibit significantly lower coercivity than their crystalline counterparts. In order to verify this statement, the magnetic properties of an Fe/Si sample and an Fe/Si + $^{11}B_4C$ sample with 14 vol.% $^{11}B_4C$ within the Fe and Si layers were studied using vibrating sample magnetometry (VSM). The results, depicted in Figure S1(b), reveal that the inclusion of 14 vol.% of $^{11}B_4C$ indeed eliminated the coercivity, and resulted in a difference of more than an order of magnitude in the external magnetic field required to saturate the magnetization. To saturate the Fe/Si at least 10 mT were needed while Fe/Si + $^{11}B_4C$ barely needed 1 mT. It Is important to note that the 14 vol.% lower Fe content in the $^{11}B_4C$ incorporated sample leads to a decrease in the total magnetization, which can be observed in the amplitude drop in the $^{11}B_4C$ incorporated sample. The slightly lower magnetization affects the m-SLD by decreasing it with the same ratio[8] but could, through our concept, easily be compensated for by increasing the non-magnetic layer SLD to achieve optimal polarization (Figure 1(b)) which was then also done as seen in Figure 4(g).

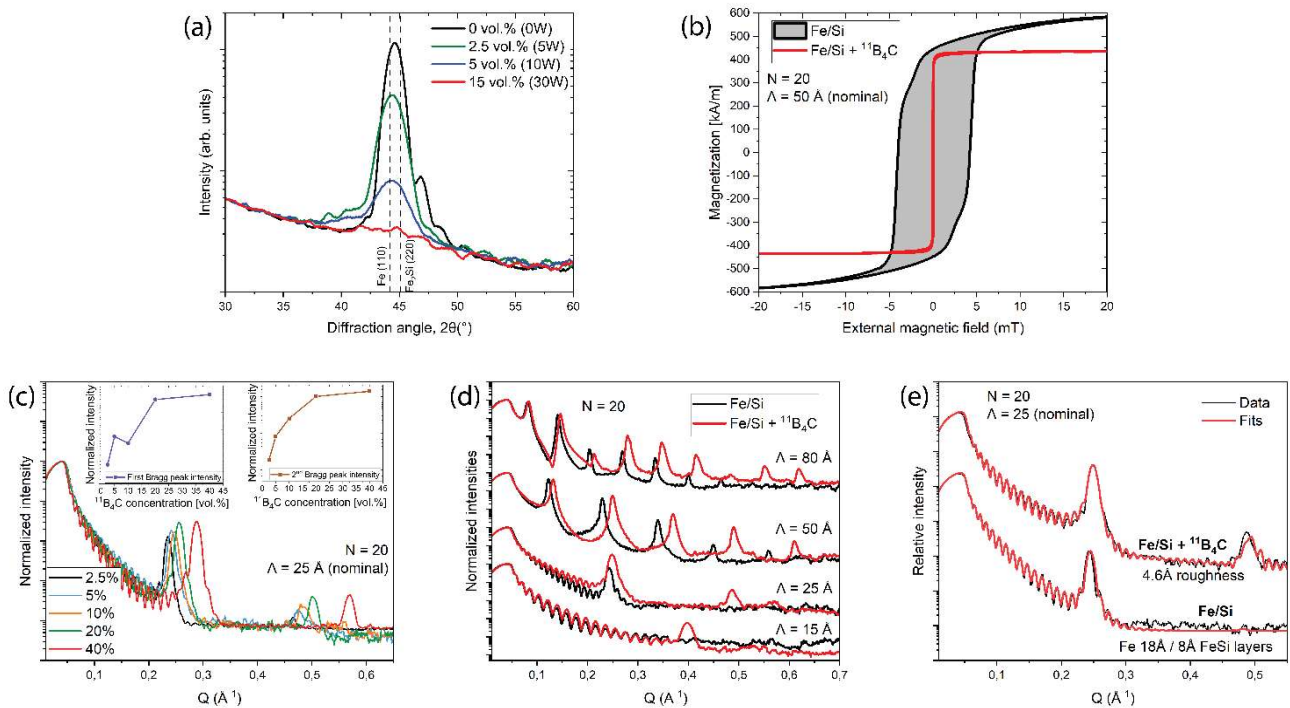

**Figure S1. Structure and magnetism.** (a) X-ray diffraction (XRD) of 4 multilayers of Fe/Si with various volumetric percentages of $^{11}B_4C$ incorporated within the layers in the multilayer, ranging from 0 vol.% to 14 vol.%. The positions for metallic Fe 110 peak and the $Fe_2Si$ silicide 220 peak are indicated. (b) Vibrating sample magnetometry (VSM) of a Fe/Si and a Fe/Si + $^{11}B_4C$ multilayer showing the hysteresis curves, where the grey area highlights the area within the magnetic coercivity. (c) X-ray reflectivity of Fe/Si + $^{11}B_4C$ samples with various ratios of $^{11}B_4C$ within the multilayers. The insets show the first (left) and second (right) order Bragg peak intensity as a function over the concentration of $^{11}B_4C$ within the multilayers. (d) X-ray reflectivity comparisons of Fe/Si and Fe/Si + $^{11}B_4C$ multilayers of various bilayer thicknesses, ranging from 15Å to 80 Å bilayer thickness. The curves have been stacked vertically for a clearer comparison between the material system (e) X-ray reflectivity data and fits of a Fe/Si and Fe/Si + $^{11}B_4C$ sample.

The effect on the reflective properties for various concentrations of $^{11}B_4C$ within the Fe/Si multilayers can be seen in S1(c) where the Bragg peak intensity increases with the increasing concentration of $^{11}B_4C$ within the multilayer. The intention was to try to keep the same bilayer thickness for comparison, so the higher the $^{11}B_4C$ content the lower the Fe and Si content. Although the nominal bilayer thickness was 25 Å, with increasing $^{11}B_4C$ concentration the bilayer decreased in thickness, which is evident from the shift of the Bragg peak to higher scattering angles. However, this further solidifies the fact that the interface width decreased with higher amounts of $^{11}B_4C$ since the Bragg peak increased even though the intensity should decrease due to the Bragg peak appearing at higher scattering angles, according to Porod's law. Further, the addition of $^{11}B_4C$ in both layers decreases the X-ray SLD contrast, hence should in theory decrease the intensity of the Bragg peak with increasing concentrations of $^{11}B_4C$. However, apparent in S1(c), the Bragg peak still increased in intensity with higher concentrations, presenting evidence that the interface width decreased. However, too high concentrations of $^{11}B_4C$ could reduce the neutron polarization capabilities due to a decrease in magnetization. Thus 14 vol.% within the Fe layers was chosen to maintain sufficient magnetization while still being magnetically soft for the continuation of our study. Observing S1(d), the incorporation of $^{11}B_4C$ resulted in more intense Bragg peaks and higher orders of Bragg peaks for all measured bilayer thicknesses, supporting our idea of decreased interface width, regardless of bilayer thicknesses. The presence of Bragg peaks at higher Q-values enables the possibility of reflectivity at higher Q-regions. The figure also illustrates a Bragg peak for a bilayer thickness of 15 Å when $^{11}B_4C$ is present, whereas no Bragg peak is observed without $^{11}B_4C$. GenX3[21] was used to fit the Fe/Si and Fe/Si + $^{11}B_4C$ samples with 25 Å nominal bilayer thickness, as shown in S1(e). The analysis suggests that the Fe/Si sample contains 26 Å of heavily mixed Fe and Si atoms throughout the entire sample which suggests iron-silicides throughout the entire multilayer, however with approximately 8 Å Si rich areas resulting in the Bragg peak that we see in S1(e). The addition of $^{11}B_4C$ prevented the formation of iron-silicide, resulting in a more clearly separated magnetic and non-magnetic layer, a decreased interface width, and increased reflectivity, along with an additional order Bragg peak.

## Section S2. XPS and chemical bonding

Both Fe *2p* and Si 2p peaks shift towards higher binding energy for the $^{11}B_4C$ containing sample as compared to the Fe/Si sample. The shift is 0.65 eV and 0.45 eV, respectively. This can be explained by the formation of Fe-B, Fe-C, Si-B, and Si-C bonds. As both B and C are more electronegative than Fe and Si, the valence charge density on Fe and Si atoms is expected to decrease upon bonding. Thus, the hypothesis that the Fe-B bonds are what causes the amorphization is then confirmed.

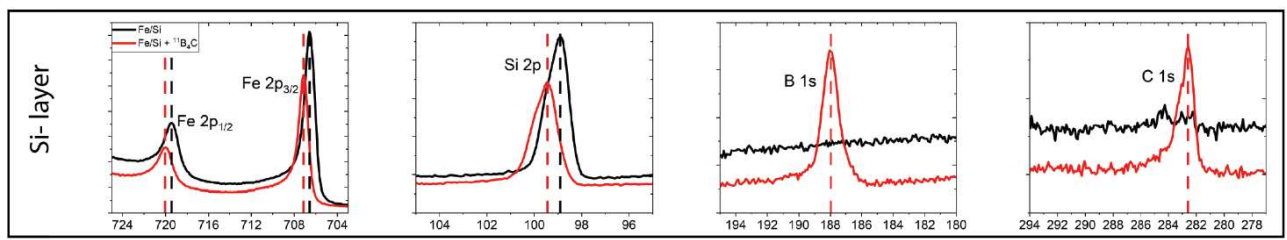

**Figure S2.** XPS spectra from the Fe/Si (red) and Fe/Si + $^{11}B_4C$ (black) multilayer (N = 10, Γ = 0.5 (nominal) and Λ = 100 Å (nominal)), acquired within the top Si layer. The spectra from left to right Fe *2p*, Si *2p*, B *1s* and C *1s* respectively.

## Section S3. TEM and HAADF-STEM with EELS

From Figure S3 it seems like the Fe/Si + $^{11}B_4C$ multilayer does not have an even distribution of B throughout the multilayer or within each layer. There seems to be a B-poor region within the Si layers. Combined with the results from the XPS, in S2, it is suggested that due to Si-C bonds within the Si layer the B would have a smaller probability to be deposited in Si rich regions.

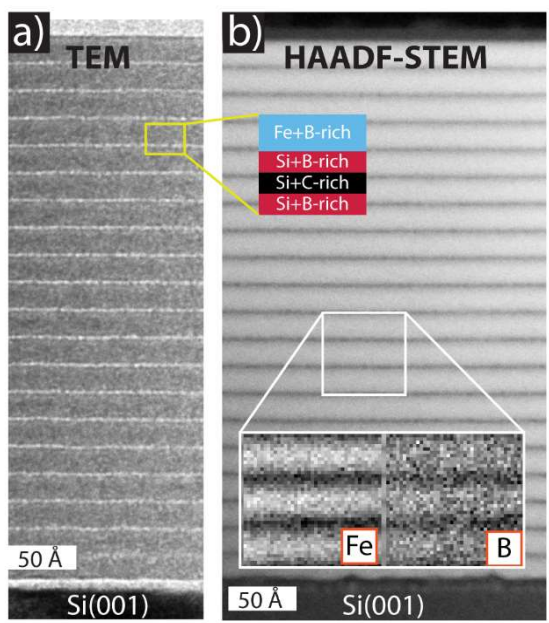

**Figure S3** (a) Section of overview diffraction contrast TEM micrograph of Fe/Si + $^{11}B_4C$ multilayer with 14 vol.% of $^{11}B_4C$ from Figure 4(d-f) and (b) corresponding scanning HAADF-STEM image and overlayed 150 x 150 Å$^2$ electron energy loss spectroscopy (EELS) map of Fe and B. From elemental analysis obtained using XPS, TEM, STEM, and EELS a schematic illustration of the elemental distribution in the multilayer stack is shown in the image.

## Section S4. GIWAXS

Figure S4(a-b) shows the grazing-incidence wide angle X-ray scattering (GIWAXS) of Fe/Si and Fe/Si + $^{11}B_4C$ multilayers, where (a) shows multilayers with a bilayer thickness of 100 Å and N = 10 and (b) multilayers with a bilayer thickness of 25 Å and N = 20. The GIWAXS data was taken simultaneously as the GISAXS measurements at the MiNaXS/P03 beamline using a LAMBDA 9M detector (X-Spectrum, pixel size = 55 µm) at a sample-to-detector distance of 284 mm. The background subtracted data is plotted using a temperature scale for intensity.

Grazing-incidence small-angle X-ray scattering (GIWAXS), Figures S4(a-b), shows the crystallinity of Fe/Si multilayers with various concentrations of $^{11}B_4C$. The quarter circles and ordered intensities stem from crystallinity within the samples known as Debye-Scherrer rings.

The intensity in the Fe/Si sample is concentrated on specific spots on the quarter circle, indicating a more defined ordering compared to a sample where the intensity is evenly distributed over the quarter circle. The broadness of the ring is related to the size of the crystallites. On the other hand, the GIWAXS pattern of the Fe/Si + $^{11}B_4C$ sample shows a broad and featureless subtle ring, indicating a lack of well-defined crystallographic planes in the material. However, it is interesting to note that the Fe/Si + $^{11}B_4C$ sample still retains this hint of the quarter circle seen in the Fe/Si sample. By adjusting the colormap intensity, it was possible to investigate a series of Fe/Si multilayers with varying concentrations of $^{11}B_4C$ to examine the transition from crystalline to amorphous in detail, as seen in Fig. S4(b). The intensity of the quarter circle shows a clear trend of going from well-defined and intense crystallinity to non-existent with increasing $^{11}B_4C$ concentration up to 40 vol.%. However, traces of crystallinity were detected up to 20 vol.% when compared to XRD results. This indicates that GIWAXS, with its higher sensitivity to crystallinity, was able to detect even small traces of crystallinity for higher $^{11}B_4C$ concentrations. Overall, the results demonstrate the effectiveness of GIWAXS in identifying the presence of crystallinity in multilayer samples with varying concentrations of $^{11}B_4C$.

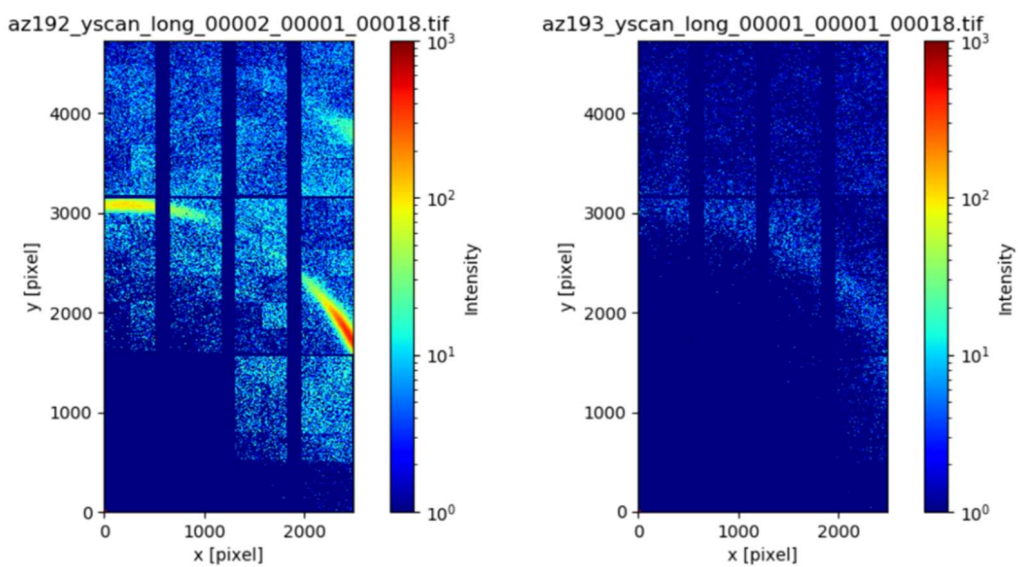

**Figure S4(a).** Grazing-incidence wide angle X-ray scattering (GIWAXS) patterns of Fe/Si (left) and Fe/Si + $^{11}B_4C$ (right) multilayers with a bilayer thickness of 100 Å and N = 10. The Fe/Si + $^{11}B_4C$ multilayer with a concentration of 14 vol.% of $^{11}B_4C$.

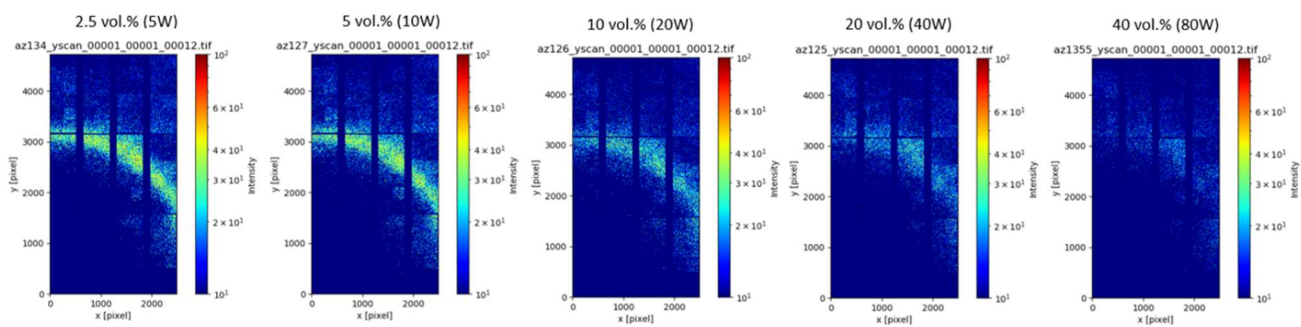

**Figure S4(b).** GIWAXS patterns of Fe/Si multilayers with varying concentrations of $^{11}B_4C$, with a fixed bilayer thickness of 25 Å and N = 20. Samples grown in a different chamber, see S7.

**Section S5. Simulations of thicker bilayers using the extracted fitted parameters from Figure 2.**

Figure S5 shows the simulated polarizing neutron reflectivity using the fitted parameters from Figure 2(b). S5(a) represents the Fe/Si multilayer while (b) displays the Fe/Si + $^{11}B_4C$. It should be noted that the polarization results stated in the main text is related to the multilayers with the specified bilayer thicknesses and number of bilayers. For thicker bilayers, however, e.g. 200 Å, the polarization values are here predicted to not be significantly increased unless finer SLD tuning is performed.

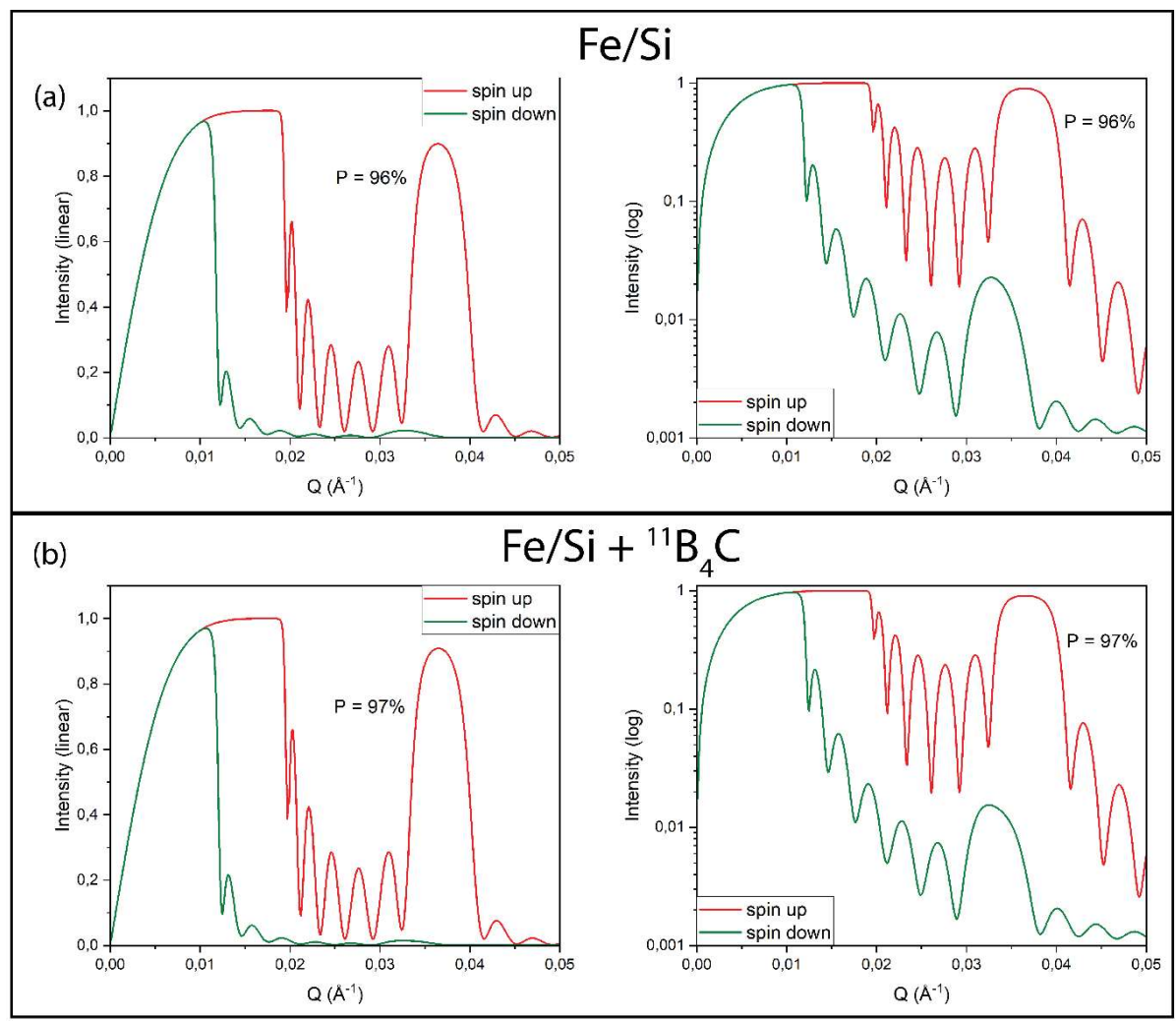

**Figure S5. Polarized neutron reflectivity (PNR) simulations using the fitted parameters from Figure 2(b). Linear intensity scale on left side and logarithmic on the right.** (a) PNR for Fe/Si and (b) PNR for Fe/Si + $^{11}B_4C$ multilayers, both with $\Lambda$ = 200 Å and N = 7.

Using the fitted parameters obtained from Figure 2(b), thicker bilayers were simulated, demonstrating that the polarization typically approaches 100%. Our study aimed to reflect and polarize at scattering angles/vectors that are not yet utilized or achievable by state-of-the-art neutron optics. Incorporating $^{11}B_4C$ into Fe/Si multilayers has been simulated to increase the polarization also for thicker bilayers as well although not as significantly as for thin bilayers where SLD matching is critical. The increase in polarization for a bilayer thickness of 200 Å was only 1%. Furthermore, it is theoretically possible to finetune the concentration of $^{11}B_4C$ to achieve even higher polarization, but this requires smaller increments of $^{11}B_4C$ to find the optimal amount for polarization enhancement.

## Section S6. Elastic Recoil Detection Analysis (ERDA)

If the vol.% is a less desired measure, atomic percentages for all the samples obtained from elastic recoil detection analysis (ERDA) are presented in the tables below. Due to the depth resolution of ERDA being several nanometers and the samples of interest having layer thicknesses as small as a nanometer the following method was used to obtain the atomic percentages. All the samples measured with ERDA have the same amount of $^{11}B_4C$ in the Fe layer but various amounts in the Si layer. By measuring the amount of $^{11}B_4C$ in the whole multilayer of the Fe/Si + $^{11}B_4C$ (13.9 vol.%) sample and a sample with twice the amount of $^{11}B_4C$ in the Si layer but the same amount of $^{11}B_4C$ in the Fe layer the amount atomic pecentage of $^{11}B_4C$ within the Fe layer could then be calculated and using extrapolation also what percentages of $^{11}B_4C$ in Si the other samples should have. ERDA measurements on those other samples confirmed that the extrapolation were accurate with less than a percent's margin. Important to note however is that the atomic percentages within each layer is based on the volumetric percentages which in turn may not be very accurate as seen through XRR fits and TEM, due to the layer thicknesses differing from the nominal values.

ERDA measures the amount of each atom and since $^{11}B$ and C are not stoichiometrically 4:1 when deposited in films the atomic percentages show the total amount of $^{11}B$ atoms + C atoms. Another reason to count $^{11}B$ and C together is due to the fact that $^{11}B$ and C intensities appear too overlapped in the ToF-ERDA spectra to be able to separate them well. Although an approximate separation was made and can be seen in Table 3.

Tables 1-3 shows the atomic compositions from ERDA of all multilayers. From Table 1 together with S1(a) we can conclude that by replacing every fifth Fe atom with a $^{11}B$ (or C) atom the Fe layer becomes X-ray amorphous and magnetically soft. From Table 2 along with Figure 4(c) we obtain the atomic percentages of $^{11}B$ + C needed to achieve optimal polarization for 25 Å bilayer thickness and a thickness ratio of 0.5. However, since the thickness ratio turned out not to be 0.5 the exact percentages may not be very accurate, although the relation between the samples are. From Table 3, it can be concluded that the actual composition of the incorporated $^{11}B_4C$ is closer to $^{11}B_{\sim 5}C$.

**Table 1.** Atomic percentages of $^{11}B$ + C within the Fe layer for the various samples seen from the main text, determined using ERDA.

| Sample | $^{11}B$ + C atomic percent within Fe layer |
|---|---|
| All samples (averaged) | 20 at.% |

**Table 2.** Atomic percentages of $^{11}B$ + C within the Si layer for the various samples seen in Figure 4, determined using ERDA.

| Sample | $^{11}B$ + C atomic percent within Si layer |
|---|---|
| Fe/Si | 0.3 at.% |
| Fe/Si + $^{11}B_4C$ (13.9 vol.%) | 15 at.% |
| Fe/Si + $^{11}B_4C$ (17.4 vol.%) | 17 at.% |
| Fe/Si + $^{11}B_4C$ (24.0 vol.%) | 19 at.% |
| Fe/Si + $^{11}B_4C$ (38.7 vol.%) | 26 at.% |

**Table 3.** Atomic percentage ratio between $^{11}B$ and C within throughout the entire multilayer for the various samples seen in Figure 4, determined using ERDA.

| Sample | $^{11}B$/C ratio throughout the entire multilayer |
|---|---|
| Fe/Si | 0.3 |
| Fe/Si + $^{11}B_4C$ (13.9 vol.%) | 4.9 |
| Fe/Si + $^{11}B_4C$ (17.4 vol.%) | 4.7 |
| Fe/Si + $^{11}B_4C$ (24.0 vol.%) | 4.9 |
| Fe/Si + $^{11}B_4C$ (38.7 vol.%) | 4.7 |

**Section S7. Experimental details for deposition system for samples from S1, S3 and S4(b)**

Fe/Si and Fe/Si + $^{11}B_4C$ multilayer thin films were deposited using ion-assisted magnetron sputter deposition in a high vacuum system with a background pressure of about $5.6 \cdot 10^{-5}$ Pa ($4.2 \cdot 10^{-7}$ Torr). The multilayers were deposited onto 001-oriented single crystalline Si substrates, $10 \times 10 \times 1$ mm$^3$ in size, with a native oxide. During deposition the substrate temperature was maintained at an ambient temperature (293 K), and to improve the thickness uniformity, the substrate was spinning at 8 rpm. The substrate table was electrically isolated, enabling a substrate bias voltage to be applied in order to attract sputter gas ions from the plasma. The main difference growing these samples compared to the other sample in this article was by using a modulated ion assistance regime during the deposition.[27] Ion-assisted deposition was employed by attracting Ar-ions from the sputter plasma through a negative substrate bias of -30 V. A magnetic field, colinear with the substrate normal, was used to condense the plasma towards the substrate, thereby increasing the Ar-ion flux at the substrate. The sputtering was modulated by alternating between 0 V substrate bias for the first approximately 3 Å to then employ the -30 V substrate bias for the rest of the layer. The modulated ion-assistance scheme is for the decrease in intermixing. The sputtering targets used were Fe (99.95% purity, 75 mm diameter), $^{11}B_4C$ (99.8% chemical purity, isotopic purity >90%, 50 mm diameter), and Si (99.95% purity, 75 mm diameter). The magnetrons were continuously running during deposition and the material fluxes were controlled using computer-controlled shutters placed in front of the magnetrons for each target material. This allowed for deposition of multilayers from the separate target materials, as well as the possibility of alloying two target materials to achieve a desired composition through co-sputtering. When depositing Fe/Si + $^{11}B_4C$, each bilayer consisted of co-sputtered $^{11}B_4C$ with Fe, followed by the deposition of $^{11}B_4C$ with Si. The deposition rates of both Fe + $^{11}B_4C$ and Si + $^{11}B_4C$ were nearly equal, approximately 0.5 Å/s. When preparing samples with different ratios of Si to $^{11}B_4C$, only the deposition rate of Si was adjusted by tuning its target power.